\documentclass[twocolumn]{paper}
\usepackage[ansinew]{inputenc}
\usepackage{indentfirst}
\usepackage{amsmath}
\usepackage{amssymb}
\usepackage{graphicx}
\usepackage[squaren]{SIunits}
\usepackage{booktabs}
\usepackage{ifthen}

\title{GPRD, A Database for the Spectral Properties of Diatomic Molecules of Atmospheric Interest}

\author{P. Passarinho, M. Lino da Silva}

\institution{Centro de Física de Plasmas, Instituto Superior
Técnico, Av. Rovisco Pais, 1049--001, Lisboa, Portugal\\%
\texttt{mlinodasilva@mail.ist.utl.pt}/Fax: +351 21 841 90 13}

\begin{document}

\onecolumn

\maketitle

For some years, spectroscopic data from diatomic molecules commonly found in low-pressure plasmas and planetary atmospheres has been compiled in order to improve the knowledge of the spectral properties of such mixtures. Up to very recently, the only available reference was provided by Huber and Herzberg's classic book on the constants of diatomic molecules \cite{Huber:1979}. Since then, two different Web-based databases, DiRef \cite{DiRef} and RADEN \cite{RADEN}, provided compilations of references associated with the spectra of diatomic molecules, the latter already providing a limited amount of numerical data.

The GPRD (Gas and Plasma Radiation Database) database moves a step further, as it provides factual data -- level-dependent spectroscopic constants, Klein--Dunham coefficients, Franck--Condon factors and r-centroids, electronic and vibronic transition moments, oscillator absorption strengths and Einstein coefficients, level-dependent and overall cross-sections for continuum transitions -- and its associated references, allowing the user to simulate the relevant rovibronic transitions of the database diatomic molecules.

Although this database provides the scientific community with a novel and yet unique tool for retrieving the overall spectral properties of diatomic molecules rovibronic transitions, otherwise missing spectroscopic data is also provided for the simulation of atomic continuum radiation. The database is also linked to the most preeminent databases on atomic radiation (providing data on discrete and photoionization radiation processes) \cite{NIST,TOPbase}. Finally, vibrationally-dependent spectroscopic data for the simulation of discrete and continuum transitions is also provided for small polyatomic molecules (this is currently limited to the CO$_{2}$ molecule). The proposed data can be considered as complementary of data provided by the HITRAN \cite{HITRAN} database, regarding the simulation of the spectral properties for high-temperature gases and plasmas.

The GPRD database can be accessed online at the internet address \emph{http://cfp.ist.utl.pt/radiation/}. This database is freely available to the overall scientific community, and only requires electronic registration in order to provide full access to the overall web-based data. The database follows a tree structure in which the user firstly chooses among atomic and molecular data, then selects the specific chemical species, and finally selects the type of required data (rovibronic levels, bound-bound, bound-free or free-free transitions calculations). The author has finally to chose which spectroscopic parameters he wants to retrieve (Klein--Dunham, Einstein coefficients, etc... ), and is provided with a list of the available sets. The present version of the database encompasses over 300 different datasets and is ever growing.

The engine used in the database has been developed using the SQL language, and the user interface is written in PHP. The spectroscopic data is provided in two different formats. The first one is a spreadsheet in the XML format, which can be opened with popular software such as Excel or OpenOffice. A XSL/CSS stylesheet has been added in order to allow online viewing of the spectral data. A simplified comma-separated (.csv) text file containing the spectroscopic data is also provided. The database program is currently being run on an Apache server.

Future updates of the database will include critical comments on the available spectroscopic data by the database managers. Comments from the general spectroscopy community will also be accepted, in order to allow users to select the most relevant and accurate spectroscopic data for their own needs. Moreover, a specific section of the database, featuring experimental spectra from atomic and molecular systems, is scheduled for development. This will allow critical comparisons between synthetic spectra (generated from the different available datasets) and measured spectra.

\subsection*{Acknowledgements}
\footnotesize%

The authors would like to acknowledge F. Passarinho for his assistance and contributions in designing the page layout.

\end{document}